\documentclass[runningheads]{svmult}

\usepackage{makeidx}   % allows index generation
\usepackage{graphicx}  % standard LaTeX graphics tool
\usepackage{subeqnar}  % subnumbers individual equations
\usepackage{multicol}  % used for the two-column index
\usepackage{physprbb}  % modified textarea for proceedings,
\makeindex             % used for the subject index
\begin{document}
\title*{Extragalactic science at very high spectral resolution}
%
%
% allows explicit linebreak for the table of content
%
%
\titlerunning{Galaxies at high spectral resolution}
% allows abbreviation of title, if the full title is too long
% to fit in the running head
%
\author{Eric Emsellem\inst{1}}
\authorrunning{Eric Emsellem}
% if there are more than two authors,
% please abbreviate author list for running head
%
%
\institute{CRAL, Observatoire de Lyon, 9 av. Charles Andr\'e, 69561 
	Saint Genis Laval, France}

\maketitle              % typesets the title of the contribution

\begin{abstract}
I briefly mention a few possible applications of very high spectral resolution
spectroscopy with CRIRES to the study of nearby galaxies. This includes the
fields of AGN, dynamically cold systems, super stellar and emission line 
clusters, and a more speculative program on the measurement of gravitational redshifts. 
\end{abstract}

\section{Introduction}
One of the early attempt to use very high spectral resolution spectroscopy to
study the central region of a nearby galaxy was motivated by the complex nature
of the observed source: Pelat \& Alloin (1980) used the spectral resolution
delivered by an Echelle spectrograph at the Observatoire de Haute Provence
as a compromise to compensate for the lack of spatial resolution. The intricate velocity
structure of the [OIII] line in the central 200~pc of the active galaxy
NGC~1068 was thus revealed, but even at a resolution of about 20,000, 
the lack of spatial resolution prevented any truly conclusive interpretation.

\begin{figure}[ht]
\begin{center}
\includegraphics[width=.45\textwidth]{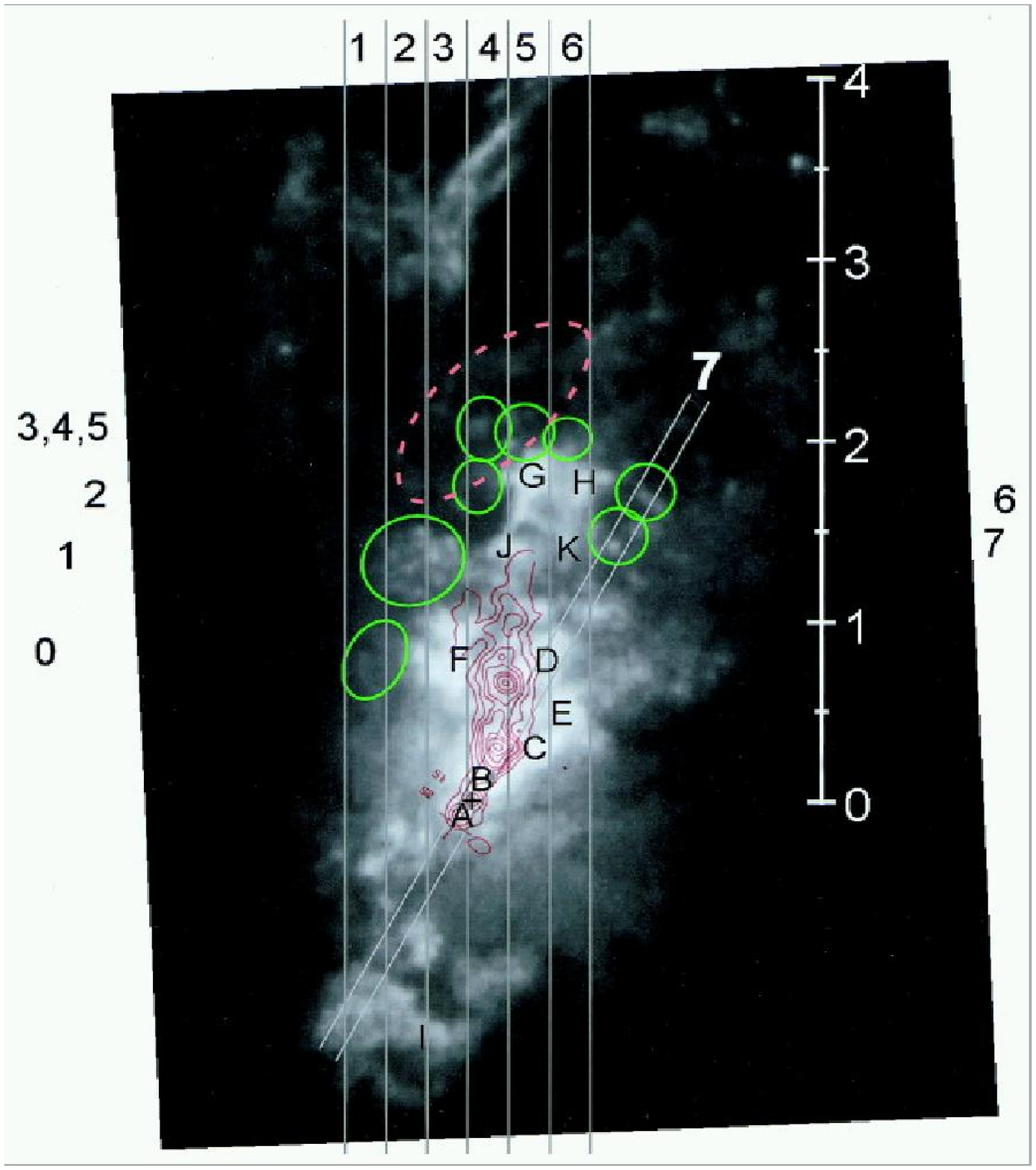}
\includegraphics[width=.54\textwidth]{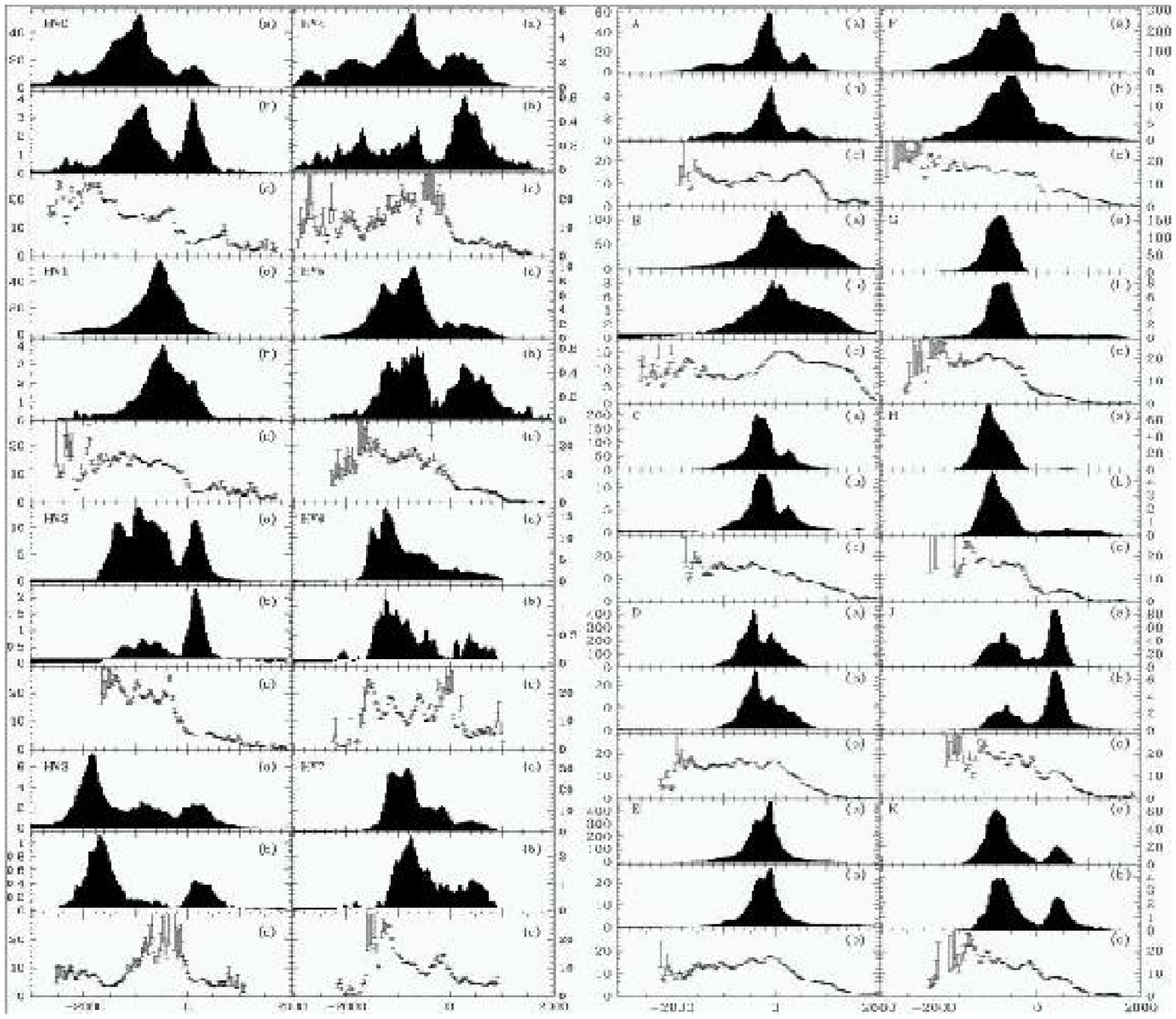}
\end{center}
\caption[]{{\em Left panel:} FOC image of [OIII] 4959, 5007 line emission across
	the circumnuclear region of NGC 1068 (Macchetto et al. 1994). The gray scale
	is proportional to the log of the intensity.  {\em Right panel:}
	Dereddened line profiles of some high-velocity knots and more massive clouds.
	In each set of three spectra, the top profile (a) is [OIII] 5007 after
	contamination by 4959 has been removed, the middle (b) is H$\beta$, and
	the bottom (c) shows the range of the ratio [OIII] 5007/H$\beta$ at each
	velocity (spectral resolution of $60$~km.s$^{-1}$ FWHM). Figures extracted
	from Cecil et al.  (2002).}
\label{fig:n1068a}
\end{figure}

A new step was recently taken with the high spatial resolution provided by the Hubble
Space Telescope (HST). Cecil et al. (2002) used STIS aboard HST to map the
central region of NGC~1068 in the [OIII] and H$\beta$ lines with 6 parallel
slits, with the aim of resolving individual clouds and their kinematics. Even at
a resolution of about 5000, the data clearly showed that there is no hope to
nail down the velocity distribution in this region.  As shown in
Fig.~\ref{fig:n1068a},  each spatial pixel includes numerous
velocity components which themselves correspond to many unresolved clouds
with various physical conditions. This illustrates the
real potential of an instrument such as CRIRES, which will simultaneously 
deliver very high spectral resolution (up to $R=100,000$) 
and high spatial resolution. 

In the following, I provide a short and non exhaustive list
of scientific programs which could benefit from such a unique spectrograph.
However I should first emphasize that, in general, CRIRES may not be the best
adapted instrument to conduct detailed studies on nearby galaxies. Indeed, 
its maximum spectral resolving power of 100,000 requires bright sources.
For most (nearby) extragalactic sources, the available flux is just not high
enough. There are still a few niches for CRIRES which should be exploited, some
of which I mention below.

\section{Diagnostics}

The Near Infrared domain available with CRIRES (1--5 $\mu$m) includes many
lines to study the physics of both the gas components and the stellar
populations. A number of atomic and molecular emission lines can be observed, e.g. 
when the region under scrutiny is highly extincted, Br$\beta$ and Br$\alpha$ could really help,
and the Pfund series could serve as a nice tool to get some handle on $A_V$.
H$_2$ lines are often present, as well as higher excitation lines such as [FeII], [SiVI],
..., thus delivering excellent diagnostics of the physical status of the gas.
The various line ratios would then inform us about the physical emission
process: collisional excitation, ultraviolet fluorescence, X ray heating, fast
shocks. 
\begin{figure}[ht]
\begin{center}
\includegraphics[width=0.7\textwidth,angle=-90]{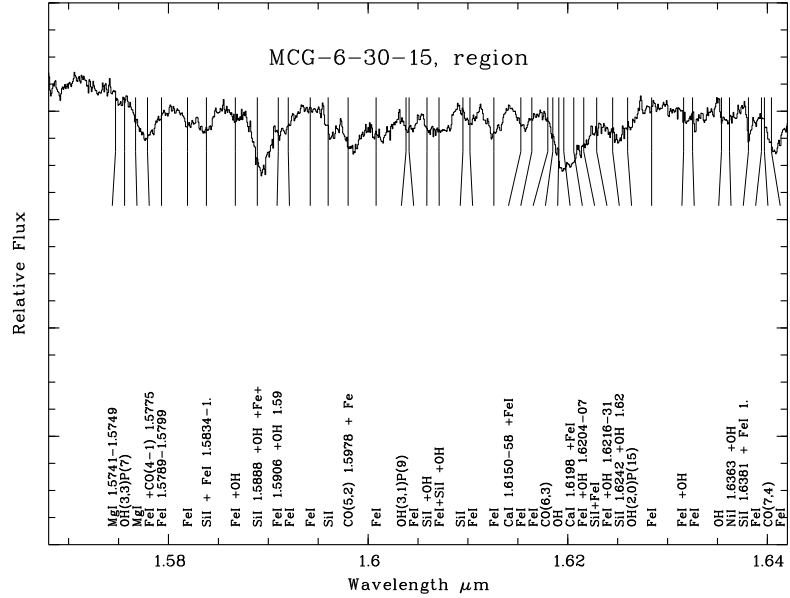}
\end{center}
\caption[]{Absorption line identification in a small part of the NIR H band.
(Courtesy of C. Boisson)}
\label{fig:absline}
\end{figure}

The stellar component is easily revealed by the numerous absorption lines which
are present at these wavelengths: e.g., FeI, TiI, MgI lines, CO bandheads at 1.6 and
2.3~$\mu$m (Fig.~\ref{fig:absline}; courtesy C. Boisson).
These are powerful tools to derive the stellar kinematics and constrain the
characteristics (age, metallicity) of the underlying stellar populations.
However this will require to have reference template spectra at the same very
high spectral resolution. This therefore implies a significant effort to observe
with CRIRES a large sample of stars with a wide range of type and properties.

\section{Dynamically cold systems and central stellar populations}

The high spectral resolution provided by CRIRES could certainly help studying
the kinematics and stellar populations 
of dynamically cold systems. Dwarfs and spheroidals are thus excellent
targets to constrain the radial profiles of dark matter, and we could think
also of measuring the stellar kinematics of face-on disks to reveal their internal 
dynamical structure. However these sources are just too faint for CRIRES, 
even with an aperture such as a VLT unit. 

As we go towards the lower end of the black hole mass (M$_{bh}$) versus velocity
dispersion ($\sigma$) relation, the velocity resolution required to probe e.g. the
centers of inactive galaxies increases. For late type spirals (Sc
and later), $R=30,000$  would probably be enough, with no need to reach
the maximum resolution of CRIRES. The latter may be useful to reach the
extreme end of that relation, i.e. to constrain the presence and mass of the very low
mass black holes in globular clusters, although we would there be again starving
for photons. Intermediate resolution of the order of 30,000 could also be used
to help disentangling the blended metallic lines, hence to constrain the
populations in the central regions of spirals. Spatial resolution is also a
requirement here since significantly more that half of all spirals 
show the presence of relatively young stellar nuclei with $-14< \mbox{M}_I < -9$
and a median effective radius of 3.5~pc (B\"oker et al. 2003, 2004). 

\section{AGN studies}

Although there will be only a handful of active nuclei for which sufficiently
high signal to noise ratios can be reached with CRIRES, it will deliver
unprecedented information on the physical status of the gaseous component in the
central few hundreds parcsecs of AGN. As illustrated in the introduction of this paper,
the spectral resolution will directly address the gas velocity structures.
Resolving these below the thermal width is a reachable goal using lines such as
[FeII] or Br$\gamma$ for lighthouses like NGC~1068 or Circinus. 
The very high velocity resolution achieved by CRIRES spectrography will also allow 
tracing the physical status of the gas independently for each velocity slice
through e.g. the line ratios. The integrated stellar populations could then be
analyzed in great details in the context of the AGN / nuclear star formation
debate. Last but not least, we should be ready for surprises, as CRIRES data
will definitely probe new grounds and hopefully reveal unexpected sights.

\section{Super stellar and emission line clusters}

\begin{figure}[ht]
\begin{center}
\includegraphics[width=\textwidth]{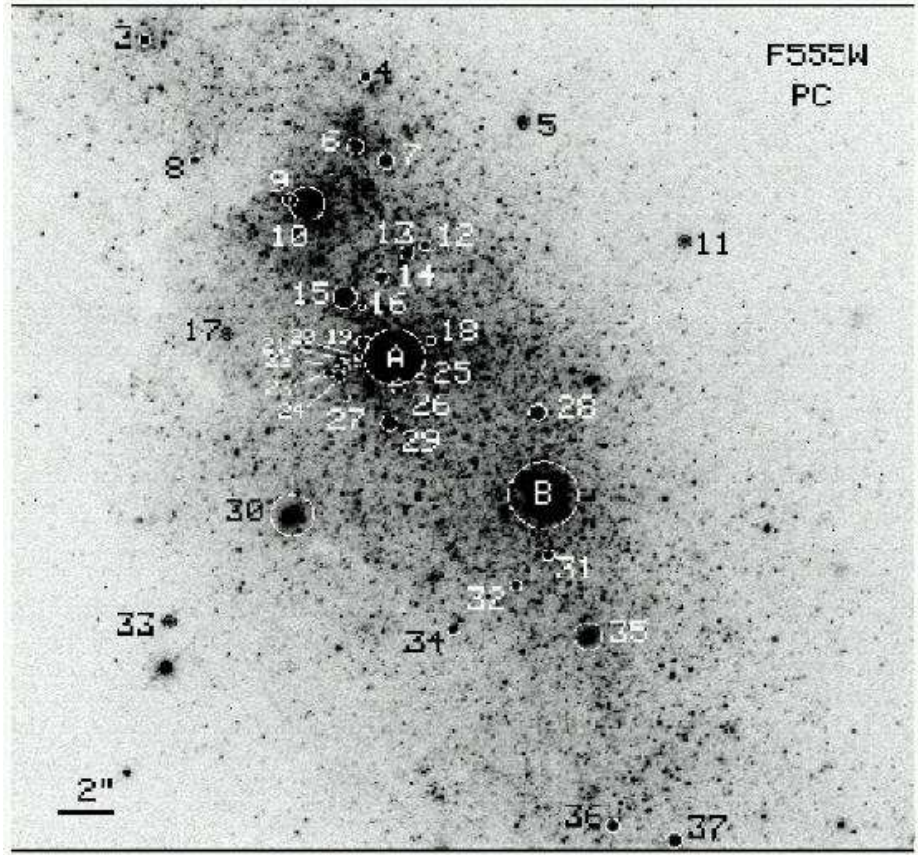}
\end{center}
\caption[]{Stars clusters observed in NGC~1569 with the WFPC2/HST camera.
Extracted from Hunter et al. (2000).}
\label{fig:ssc}
\end{figure}

Another field where CRIRES could definitely contribute is the study of super
stellar clusters in nearby galaxies. Such clusters are often observed in
starburst galaxies and correspond in size and luminosities to globular clusters
(Hunter et al. 2000). An image obtained with the WFPC2 camera aboard HST is
provided in Fig.~\ref{fig:ssc} as an illustration. The apparent size of these
clusters requires high spatial resolution, and the mass range they probe imposes
high spectral resolution. 

\begin{figure}[ht]
\begin{center}
\includegraphics[width=\textwidth]{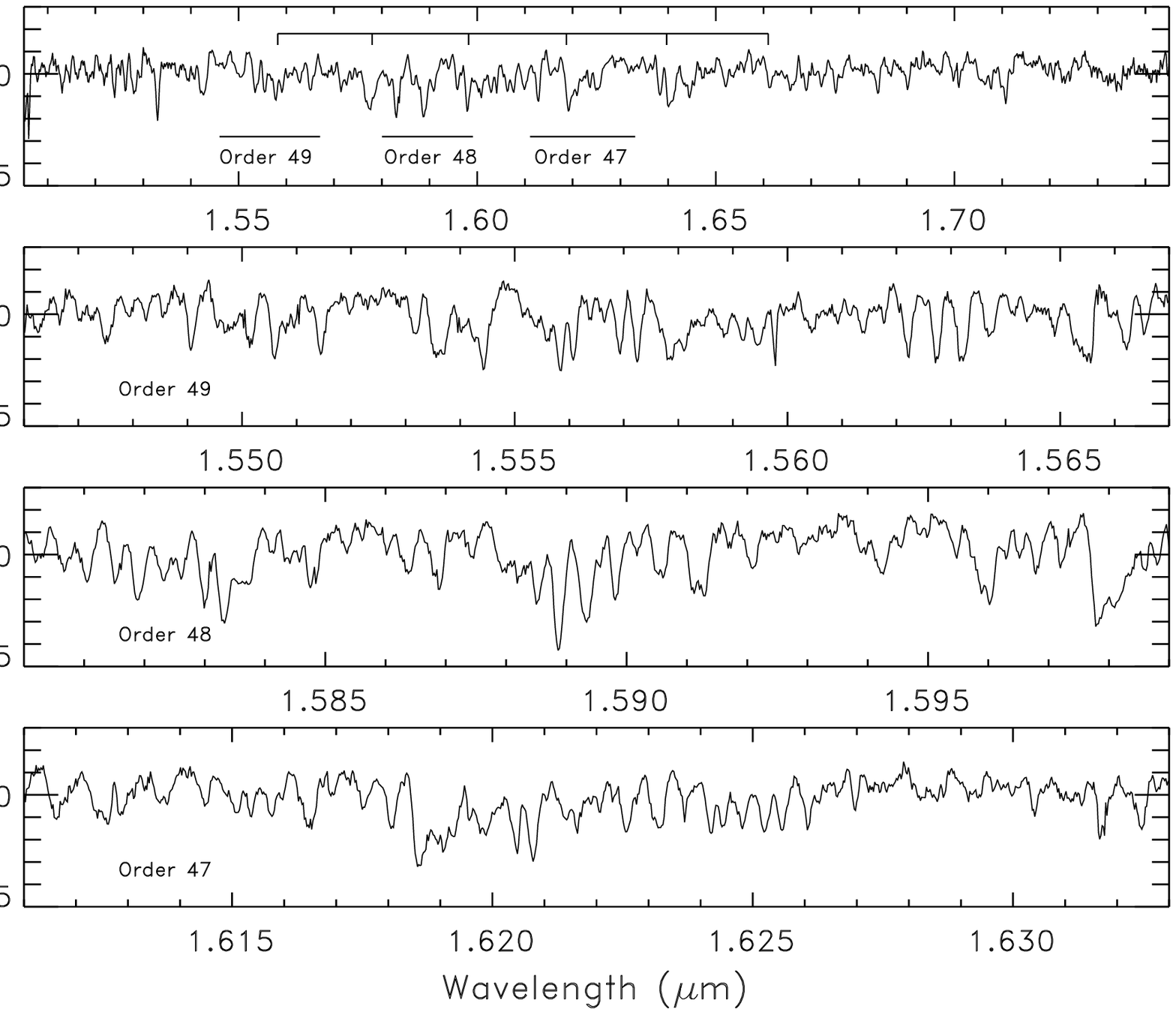}
\end{center}
\caption[]{Top 2 panels: H band spectra of a SSC B, a super stellar cluster in NGC~1569
at a resolution of $R~2000$. Bottom 2 panels: zooming with echelle spectra at $R~24,000$.
From Gilbert (2002).}
\label{fig:spec_ssc}
\end{figure}
\begin{figure}[ht]
\begin{center}
\includegraphics[width=\textwidth]{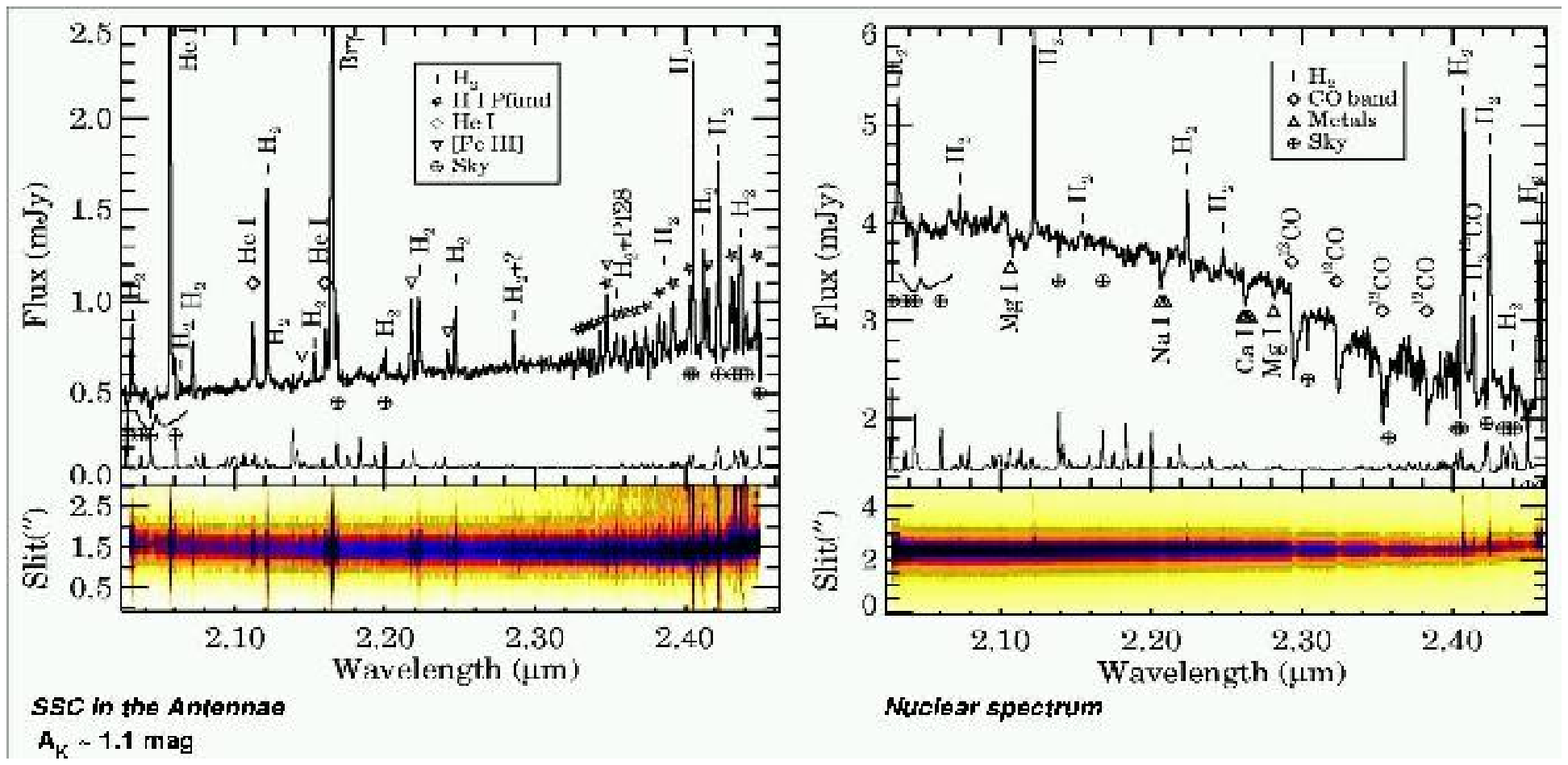}
\end{center}
\caption[]{NIRSPEC Keck II spectra of an obscure stellar cluster in the Antennae
(left panel), and of NGC~4039 nucleus (right panel). Extracted from Gilbert et al. (2000).}
\label{fig:spec_ssc2}
\end{figure}

Spectra at $R\sim 24,000$ of SSC~B in NGC~1569 (M$_H=12.3$~mag, 
Gilbert 2002) distincly show the rich absorption 
line signatures present in the NIR domain, and only properly revealed with $R >
20,000$ (Fig.~\ref{fig:spec_ssc}, courtesy of A. Gilbert, PhD Thesis, 2002). 
The masses of these clusters can be estimated via a measurement of the
velocity dispersion (Gilbert 2002). The spectral information is also
critical to constrain their age and metallicity: 
these are unique probes of the star formation processes
in this mass range (and extreme conditions). Further illustrations can also be
found in e.g. Mc Crady et al. (2003) for the case of a well-known starbursting
galaxy, M~82, and in Gilbert et al. (2000) for a study of super stellar
clusters in the Antennae galaxies. The latter study also emphasizes the richness
of both absorption and emission lines in the NIR (Fig.~\ref{fig:spec_ssc2}), 
where both stellar and (atomic plus molecular) gas contributions are present.

The youngest members of these clusters are still embedded in HII region
and heavily extincted, with A$_V$ up to 10 and more. As they may
subsequently evolve into bona-fide globular clusters, it would be important to
examine their nebular content as a probe of the early stages of massive star
forming regions. Gilbert \& Graham (2004) have recently examined such clusters
in the Antennae galaxies and discovered supersonic Br$\gamma$ lines in their
high resolution echelle spectra. With widths of up to 105~km.s$^{-1}$, the
observed emission lines correspond to the higher end of the flux vs line width
correlation known for giant extragalactic HII regions. The so-called Emission
Line Clusters (ELCs) thus exhibit massive gas outflows, and would certainly
deserves observational campaigns with high resolution spectrographs like
CRIRES.

\section{Gravitational redshifts?}

On the speculative side, high spectral resolution spectroscopy 
may also provide a unique leverage on the direct determination of the mass of
black holes in the centre of galaxies. The potential well has a direct influence
on the frequency of the escaping photons. The Doppler effect is directly
proportional to $\Delta \Phi / c$ where $\Delta \Phi$ represents the change in
gravitational potential. We could therefore try to look for the relative Doppler
shift induced by the strong potential gradient at the centre of galaxies. The
expected mean effect is of the order of a few km.s$^{-1}$ for supermassive black
holes in the centre of luminous early-type galaxies. We would then require
spectroscopy with rather high spectral {\em and} spatial resolutions. 

Although it seems illusory to disentangle this effect from non-gravitational motions in
the gaseous component at scales of 10's of parcsecs (Popovic et al. 1995), I would argue that such a study
using stellar kinematics may be viable (see Stiavelli \& Seti 1993). 
As the observed effect will be luminosity
weighted, we need to focus on galactic centers with very cuspy luminosity
densities (modest size ellipticals), but to maximize 
the potential gradient we would require objects with high velocity dispersion 
(massive ellipticals). There are of course many limitations that we can foresee:
systematics must be mastered, unresolved structures can induce artefacts, and
intrinsic redshifted populations would confuse the picture. Line Of Sight
Velocity Distributions at such high resolutions have never been obtained, so
this may be an opportunity for an exciting experiment. However, this experiment
will probably require a number of tested cases before any strong conlusions can
be made.

\section{Conclusions}

$R=100,1000$ spectroscopy is definitely too high for most of the targets we are
used to when we study nearby extragalactic objects. However, there are still
a few niches that a unique instrument like CRIRES can fill. The
simultaneous delivery of high spatial and spectral resolutions will certainly be
a productive tool to probe the central regions of active galactic nuclei. It may
also serve its purpose to constrain the central stellar populations in late type
spirals, or for the search of the lower end of central dark masses. Young
massive clusters (SSCs and ELCs) are exciting targets to examine with CRIRES,
thus entering the early stages of massive star forming regions. I then
suggested that high spectral resolution spectroscopy could be a way to get a
direct measurement of the potential well via gravitational redshifts
detections. Finally, and although it seems clear from this short paper that
extragalactic science will not be a privileged field for CRIRES, we should keep
our minds open for surprises: CRIRES will certainly drive us towards unknown
territories. 

I would like to take this opportunity to 
thank the organizers and particularly Alan Morwood
for the invitation to this exciting meeting.
I also wish to sincerely thank Andrea
Gilbert, Eva Schinnerer and Catherine Boisson, for insightful discussions, 
and for sharing some of their unpublished results.


\begin{thebibliography}{11.}
\addcontentsline{toc}{section}{References}
\bibitem{2003SPIE.4834...57B} B{\" o}ker, T., van der 
Marel, R.~P., Gerssen, J., Walcher, J., Rix, H., Shields, J.~C., \& Ho, 
L.~C.\ 2003, SPIE, 4834, 57 

\bibitem{2004AJ....127..105B} B{\" o}ker, T., 
Sarzi, M., McLaughlin, D.~E., van der Marel, R.~P., Rix, H., Ho, L.~C., \& 
Shields, J.~C.\ 2004, AJ, 127, 105 

\bibitem{2002ApJ...568..627C} Cecil, G., Dopita, M.~A., 
Groves, B., Wilson, A.~S., Ferruit, P., P{\' e}contal, E., \& Binette, L.\ 
2002, ApJ, 568, 627

\bibitem{2002PhDT........15G} Gilbert, A.~M.\ 2002, 
Ph.D.~Thesis, Univ. of California, Berkeley

\bibitem{2000ApJ...533L..57G} Gilbert, A.~M., et al.\ 
2000, ApJL, 533, L57 

\bibitem{2003IAUS..221E..63G} Gilbert, A.~M.~\& 
Graham, J.~R.\ 2003, IAU Symposium, 221

\bibitem{2000AJ....120.2383H} Hunter, D.~A., O'Connell, R.~W., 
Gallagher, J.~S., \& Smecker-Hane, T.~A.\ 2000, AJ, 120, 2383 

\bibitem{2003ApJ...596..240M} McCrady, 
N., Gilbert, A.~M., \& Graham, J.~R.\ 2003, ApJ, 596, 240 

\bibitem{1980A&A....81..172P} Pelat, D.~\& Alloin, 
D.\ 1980, A\&A, 81, 172 

\bibitem{1995A&A...293..309P} Popovic, L.~C., Vince, I., 
Atanackovic-Vukmanovic, O., \& Kubicela, A.\ 1995, A\&A, 293, 309 

\bibitem{1993MNRAS.262L..51S} Stiavelli, M.~\& 
Setti, G.\ 1993, MNRAS, 262, L51 

\end{thebibliography}
\end{document}